\documentclass[pra,aps,twocolumn,superscriptaddress]{revtex4-2}
\usepackage[T1]{fontenc}

\usepackage{amsmath,amssymb}
\usepackage{amsfonts}
\usepackage{epsfig,pstricks,graphicx}
\usepackage{bm}
\usepackage{physics}

\definecolor{armygreen}{rgb}{0.29, 0.33, 0.13}
\usepackage{color}
\usepackage{array}

\usepackage[normalem]{ulem} 
\usepackage{soul} 

\begin{document}

\title{No-go for device independent protocols with Tan-Walls-Collett `nonlocality of a single photon'}
 
\author{Tamoghna Das}
\author{Marcin Karczewski}
\author{Antonio Mandarino}
\author{Marcin Markiewicz}
\author{Bianka Woloncewicz}
\author{Marek \.Zukowski}
\affiliation{International Centre for Theory of Quantum Technologies, University of Gda\'nsk, 80-308 Gda\'sk, Poland}

\begin{abstract}
We investigate the 
interferometric  scheme put forward by Tan, Walls and Collett 
[Phys. Rev. Lett. {\bf 66}, 256 (1991)]
that aims to reveal Bell non-classicality of a single photon. By providing a local hidden variable model that reproduces their results, we decisively refute this claim. In particular, this means that the scheme cannot be used
in device-independent protocols.
\end{abstract}

\maketitle

\begin{figure}
	\centering
\includegraphics[width= 0.99 \columnwidth]{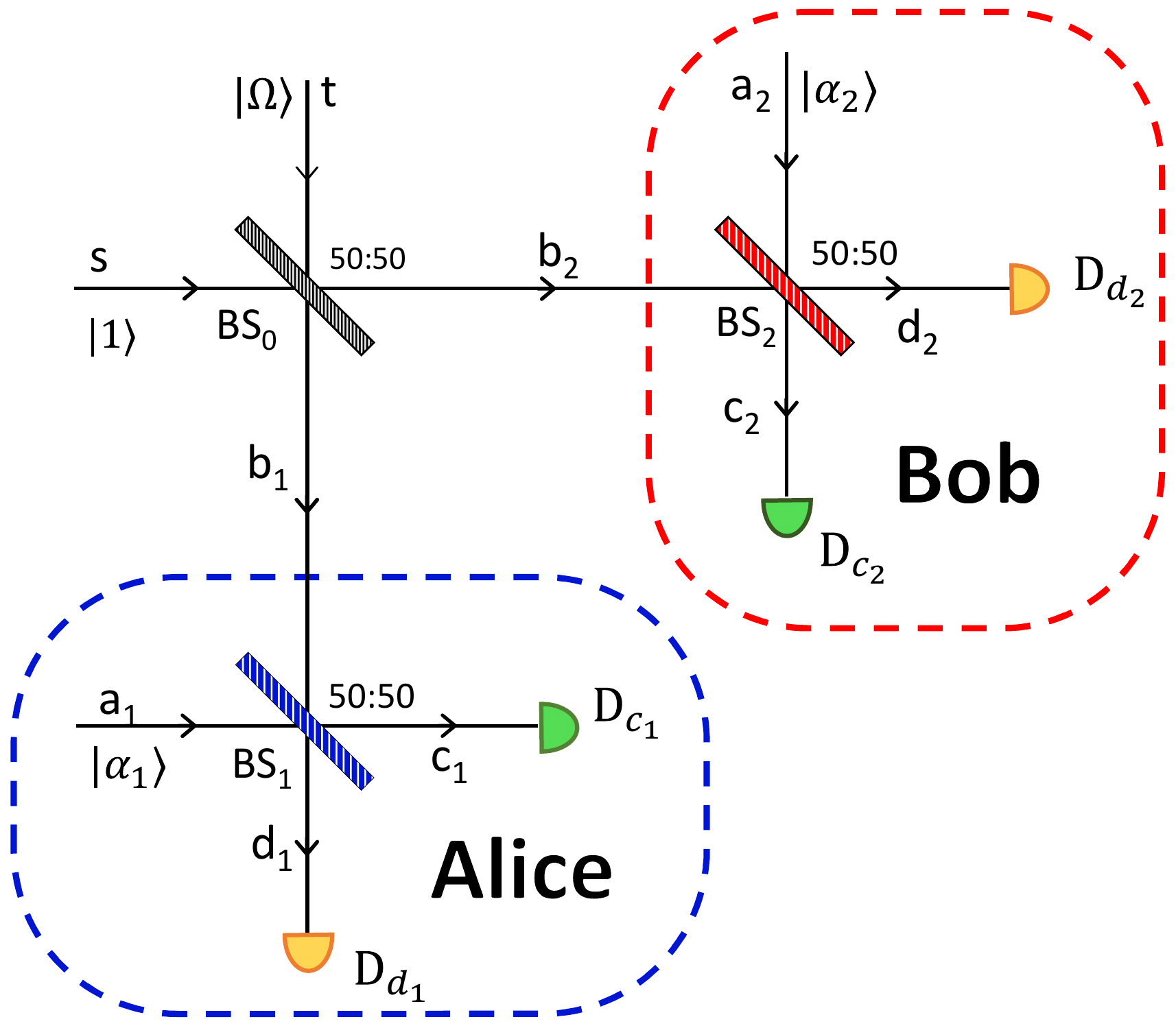}
	\caption{\label{mainSetup}
	Experimental configuration proposed  by Tan, Walls and Collett in \cite{TWC91}.  A single photon impinges on a 50-50 beamsplitter via input $s$, along with the vacuum in the input $t$. As a result we get state $\ket{\psi}_{b_1, b_2}$, which propagates to the laboratories of Alice and Bob, who perform homodyne  measurements involving weak coherent local oscillator fields (their amplitudes satisfy $|\alpha_1|=|\alpha_2|$).
	}
\end{figure}

`Nonlocality of a single photon' is a controversial and long debated subject. It was first addressed by Tan, Walls and Collett (TWC) in \cite{TWC91}. The authors aimed \emph{to demonstrate in a most striking way an effect that cannot be duplicated in any classical theory}, namely a violation of local realism with a single particle. TWC considered the state
\begin{equation}
	\label{PSI}
	\ket{\psi}_{b_1, b_2} = \frac{1}{\sqrt{2}} \left[ \ket{01}_{b_1, b_2}+ i \ket{10}_{b_1, b_2} \right],
\end{equation}
obtained by casting a single photon on a balanced beamsplitter, where e.g.
$|10\rangle_{b_1,b_2},$ indicates 
one photon excitation in the Fock space of exit mode $b_1$ and the vacuum of the Fock space relative to exit mode $b_2$, see Fig.(\ref{mainSetup}). 
The form of such state appears to be similar to the singlet state of two level systems, which is known to maximally violate a Bell's inequality. The two states are however intrinsically different in terms of the number of particles involved and $\ket{\psi}_{b_1, b_2}$ can be thought of as a plain superposition of the photon in either of the beams. 

As a result of these two opposing views, it is not intuitively obvious if it is possible to violate a Bell inequality using the single-photon state $\ket{\psi}_{b_1, b_2}$. Apart from the fundamental theoretical significance of this problem, it is also important from a practical point of view. As this state is very easy to obtain, we would like to know whether it could be treated as a resource in quantum information. In this Brief Report we will argue that its use in the device independent protocols, which rely on Bell non-classicality to work regardless of the internal functioning of the apparatus implementing them, cannot be based on the original scheme presented by TWC.

In the experimental proposal of  TWC shown in
Fig.(\ref{mainSetup}), the state $\ket{\psi}_{b_1,b_2}$ is distributed between Alice (controlling the mode $b_1$) and Bob ($b_2$).  They both perform homodyne measurements on their parts. To his end, the state in mode $b_j$ and the auxiliary coherent fields in state $\ket{\alpha_j}_{a_j}=\ket{\alpha e^{i \theta_j}}_{a_j}$ are cast into the different input ports of  50-50 beamsplitters BS$_j$, $j = 1,2$ and end up in  photon number measuring  detectors $D_{c_j}$ and $D_{d_j}$, which  register the intensities in the output ports.
 
To benchmark the violation of local realism, TWC used the
   correlation function 
   \begin{equation}
     E(\theta_1, \theta_2) = \frac{\langle (I_{c_1}(\theta_1)-I_{d_1}(\theta_1))(I_{c_2}(\theta_2)-I_{d_2}(\theta_2))\rangle_{LHV}}{\langle(I_{c_1}(\theta_1)+I_{d_1}(\theta_1))(I_{c_2}(\theta_2)+I_{d_2}(\theta_2))\rangle_{LHV}},
   \end{equation}
   
   where $I_{x_j}(\theta_j)$ is the intensity at output $x=c,d$ measured by the observer $j=1,2$ and the averaging is done over local hidden variables (LHV). Then, they considered the inequality
\begin{equation}
\label{ineq}
   |E(\theta_1, \theta_2)+E(\theta_1', \theta_2)+E(\theta_1, \theta_2')-E(\theta_1', \theta_2')|\leq 2,
\end{equation}
in which the settings are defined by the local phases $\theta_j$ and $\theta'_j$. For the amplitudes of the local oscillators satisfying $\alpha^2<\sqrt2 -1$, they observed a violation of the inequality in formula (\ref{ineq}) and concluded that the single photon state in Eq. (\ref{PSI}) is `nonlocal'. 

However, the inequality in formula (\ref{ineq}), derived in\,\cite{Reid86}, rests on the assumption that the total intensity $I_{c_j}(\theta_j, \lambda)+I_{d_j}(\theta_j, \lambda)$ registered by each observer  does {\em not} depend on $\theta_j$.
As observed in \cite{Zukowski16} and \cite{Santos92}, we cannot be certain that this condition holds. Therefore the TWC attempt to demonstrate the Bell non-classicality of the single-photon state $\ket{\psi}_{b_1, b_2}$ has to be re-examined.

Since its appearance, the TWC letter \cite{TWC91} attracted much attention, and stirred a lot of controversy.  
Santos suggested that the intensity correlations in the TWC scheme can be explained 
with local hidden variables  \cite{Santos92} and cannot be used to convincingly demonstrate non-classicality of a single photon. However, his LHV model reproduced only the correlation functions, and not the full quantum predictions concerning, for instance, registered photon numbers. Other works challenge the single-photon nature of the effect \cite{Greenberger95, Peres95}, or suggest modifications of the experiment which would allow provable violations of local realism \cite{Hardy94, Banaszek99, Enk05}. Thus far, no definite answer was given to the problem whether the TWC interference effect, which seemingly violates local realism,
admits a precise local realistic model, or not. 

Papers describing the experimental realizations of variants of this scheme  \cite{Hessmo04, Babichev04} report  violations of a Bell inequality. However these claims were presented with caution, e.g in \cite{Babichev04}, where it is effectively stated that the results are not better than for conventional Bell tests which involve the efficiency loophole.

In this Brief Report we introduce an LHV model that reproduces precisely all quantum predictions for the TWC setup. Its applicability is limited by the strength of the local oscillators, but covers the range reported in \cite{TWC91} as revealing the `nonlocality' of the single-photon state. 
This result definitely closes the case and precludes any attempt
to implement device-independent protocols relying on TWC correlations.
On the other hand, in a forthcoming publication (\emph{in preparation}) we present a modification of the TWC setup that would allow for a genuine violation of local realism. Its main idea is that, in contrast to the TWC case, the strengths of local oscillators need to depend on settings. This means that no definite initial state can be ascribed to all quantum optical fields involved in this scheme. 


{\it Explicit LHV model of TWC correlations.} Quantum predictions for the TWC setup are fully characterized by the probabilities $p(\mathbf{n})$ of events consisting in registering a specific numbers of photons in the  output modes: $\mathbf{n}=(k_{c_1},l_{d_1},r_{c_2},s_{d_2})\in\mathbb{N}^4$  (for readability, we omit the indices indicating the modes in further parts of this report). They read (see Appendix  A for the derivation) 
\begin{equation}
p(\mathbf{n})=A(\alpha,\mathbf{n})\Big[ (k\!-\!l)^2 +(r\!-\!s)^2+ 2(k\!-\!l)(r\!-\!s)\sin(\theta_{12}) \Big],
\label{eq:probability}
\end{equation} 
where $\theta_{12}=\theta_1-\theta_2$ and
\begin{equation}
    A(\alpha,\mathbf{n})=\frac{e^{-2\alpha^2}}{ k!~ l!~ r! ~ s!} \Big(\frac{\alpha^2}{2}\Big)^{k+l+r+s} \frac{1}{2\alpha^2}.
\end{equation}

These probabilities have some features, which hint at how one could reproduce them with an LHV model. First, whenever both detectors of Alice {\it or} Bob register the same number of photons, the probability does not depend on $\theta_{12}$. Let us denote the set of these events $\mathcal{N}:=\{\mathbf{n} : k=l\,\, \text{or}\,\, r=s\}.$ 
We will cover them by a family of trivial LHV submodels assigning fixed outcomes to Alice and Bob, see further.

Next, notice that all the probabilities that do depend on $\theta_{12}$ are of the form
 \begin{equation}
 \label{prob_int}
    p(\mathbf{n})=B(\alpha,\mathbf{n})\,( 1+\mathcal{V}(\mathbf{n})\sin(\theta_{12}) ),
\end{equation}
where $B(\alpha,\mathbf{n})=A(\alpha,\mathbf{n})\Big[ (k\!-\!l)^2 +(r\!-\!s)^2\Big]$ and
 \begin{equation}
   \mathcal{V}(\mathbf{n})=\frac{2(k\!-\!l)(r\!-\!s)}{ (k\!-\!l)^2 +(r\!-\!s)^2}.
\end{equation}
To reproduce them, we will adapt a model by Larsson \cite{LARSSON99}, which reproduces all the quantum predictions for a two-qubit singlet state, provided that the detection inefficiency is lower than $2/\pi$. A variant of the Larsson model is also reproducing two-qubit Franson-type interference \cite{Franson99}.

Our model $\mathcal{M}$ is a convex combination of submodels $\mathcal{M}_\mathbf{n}$, each chosen with 
probability $P(\mathcal{M}_{\mathbf{n}})$. The submodels belong to two infinite families: the trivial $\{\mathcal{M}_\mathbf{n}\}_{\mathbf{n}\in\mathcal{N}}$ and Larsson-like one $\{\mathcal{M}_\mathbf{n}\}_{\mathbf{n}\in\mathcal{\tilde{N}}}$, where $\mathcal{\tilde{N}}:=\{\mathbf{n} : k>l\,\, \text{and}\,\, r>s\}$. 
We shall focus on the latter first.

The approach that we use differs from the one of Larsson in that we do not exploit the detection loophole.  Instead, we group the probabilities that depend on the local settings $\theta_j$ with the ones that correspond to the events in which (perfect) detectors of either Alice or Bob do not register any photons. We shall denote the set of such events by $\mathcal{O}:=\{\mathbf{n}\in\mathcal N : k=l=0\,\, \text{or} \,\,r=s=0\}$.

Each Larsson-like submodel $\{\mathcal{M}_{(k,l,r,s)}\}$ is going to predict eight events resulting from applying (or not) the swaps $k\leftrightarrow l$ and $r\leftrightarrow s$ to $(0,0,r,s)$, $(k,l,0,0)$ and $(k,l,r,s)$. Notice that only one of them matches the \emph{index} $(k,l,r,s)\in\tilde{\mathcal N}$ of the model.
To construct it, we take a  uniformly distributed continuous hidden variable $\lambda\in[0,2\pi]$ and a coin toss one $x\in\{0,1\}$,  which characterize the LHV probabilities  $P^A_{\mathbf{n}}$ and $P^B_{\mathbf{n}}$, assigned to outcomes obtained by Alice and Bob in each submodel $\mathcal{M}_{\mathbf{n}}$. Their form stems from the model presented in \cite{LARSSON99}, generalizing it to the case of $0<|\mathcal{V}(n)|\leq1$.
 
Specifically, for $x=0$  Alice can register the event $(c,d)\in\{(k,l),(l,k)\}$ with probability
\begin{eqnarray}
\label{janoke}
  &\!\! \! \!\!\!\!P_{\mathbf{n}}^A(c,d|\theta_1,\lambda,0)=R_{\mathbf{n}}(c,d|\theta_1, \lambda)= \frac{1-\mathcal{V}(\mathbf{n})}{\pi}\nonumber \\ &\!\!\!\!\!\!\!+\mathcal{V}(\mathbf{n})|\sin{(\theta_1-\lambda)}| H_{}\left((c-d)\sin(\theta_1-\lambda)\right),
  \end{eqnarray}
  where $H$ is the Heaviside function.
  Otherwise, she detects no photons at all
\begin{eqnarray}
\label{ProbA00}
  P_{\mathbf{n}}^A(c=0,d=0|\theta_1,\lambda,0) = R_{\mathbf{n}}(0,0|\theta_1,  \lambda)
  \nonumber \\
  =1-\sum\limits_{(e,f)\in\{(k,l),(l,k)\}}R_{\mathbf{n}}(e,f|\theta_1, \lambda).
\end{eqnarray}
Bob detects $(c',d')\in\{(r,s),(s,r)\}$ with probabilities 
\begin{eqnarray}
\label{janoke3}
&P_{\mathbf{n}}^B(c',d'|\theta_2,\lambda,0) \nonumber \\ &=Q_{\mathbf{n}}(c',d'|\theta_2, \lambda) =H\big((c'-d')\cos(\theta_2-\lambda)\big). 
\end{eqnarray}
For $x=1$ we swap the forms of the functions defining the hidden probabilities. 
This symmetrizes the submodel, as now
$P_{\mathbf{n}}^A(c,d|\theta_1, \lambda, 1)= Q_{\mathbf{n}}(c,d|\theta_1, \lambda)$, 
and $P_{\mathbf{n}}^B(c',d'|\theta_2, \lambda, 1)= R_{\mathbf{n}}(c',d'|\theta_2, \lambda).$

Having established the local probabilities, let us now turn our attention to the joint ones. For each submodel they are given by
\begin{eqnarray} \label{MODELLO}
\!\!&P^{AB}_{\mathbf{n}}(c,d,c',d'|\theta_1, \theta_2) \!
\nonumber \\
\!\!& =\frac{1}{{4\pi}}\sum\limits_{x=0}^1 \int_{0}^{2\pi}\!\! \!\!\!d\lambda\,  P^A_{\mathbf{n}}(c,d|\theta_1, \lambda,x)\,P^B_{\mathbf{n}}(c',d'|\theta_2, \lambda,x).
\end{eqnarray}
For instance, the probability that the submodel $\mathcal{M_{\mathbf{n}}}$ predicts the event $(k,l,r,s)$ in the simplest case of  $\pi/2> \theta_1  > \theta_2>0$, $k>l$ and $r>s$, can be calculated as follows 
\begin{eqnarray}
&\!\!P^{AB}_{\mathbf{n}}(\mathbf{n}|\theta_1, \theta_2) \nonumber\\
&=\sum\limits_{x=0}^1 \int_{0}^{2\pi} \!\!\!d\lambda\,  \frac{P^A_{\mathbf{n}}(k,l|\theta_1, \lambda,x)\,P^B_{\mathbf{n}}(r,s|\theta_2, \lambda,x)}{4\pi} =\frac{1\!-\!\mathcal{V}(\mathbf{n})}{2\pi} \nonumber\\
&\!\!\!\!+\frac{2\mathcal{V}(\mathbf{n})}{4\pi}\!\int^{\theta_1}_{\frac{\pi}{2}-\theta_2}\! d \lambda  \sin{(\theta_1\!-\!\lambda)}
= \frac{1\!+\mathcal{V}(\mathbf{n})\sin(\theta_{12})}{2\pi}.
\end{eqnarray}
All other predictions of the submodel can be obtained similarly.
For  events 
$(k,l,r,s),(l,k,r,s),(k,l,s,r)$ and $(l,k,s,r)$ we get
\begin{eqnarray}
\label{interference}
\!&P^{AB}_{\mathbf{n}}(c,d,c',d'|\theta_1, \theta_2) \nonumber \\
&=\frac{1\!+\mathcal{V}(\mathbf{n})\,\text{sign}\left((c\!-\!d)(c'\!-\!d')\right)\sin(\theta_{12})}{2\pi}.
\end{eqnarray}
In the case of the $\mathcal O$-events $(0,0,r,s),(0,0,s,r),(k,l,0,0)$ and $(l,k,0,0)$, the probability is flat and reads
$
\label{zeros}
\frac{1}{4}- \frac{1}{2\pi}$, which follows directly from the normalisation condition in Eq. (\ref{ProbA00}).
Comparing Eq. (\ref{interference}) with the corresponding quantum probabilities, we see that each Larsson-like submodel $\mathcal{M}_{\mathbf{n}}$ must appear in the  full model $\mathcal{M}$ with probability
\begin{equation}
\label{probLarsson}
    P(\mathcal{M}_{\mathbf{n}})=2\pi B(\alpha,\mathbf{n}).
\end{equation}
In the Appendix B we show that our formulas for 
$ P(\mathcal{M}_{\mathbf{n}})$ lead to a properly normalized probability distribution, with the proviso described below.

The presented model definitely reproduces all probabilities which reveal interference.
However a {\em sine qua non} condition for consistency of the full model is to properly describe  also the events of the 
$\mathcal{O}$ class. The above construction ascribes all the probabilities $(\frac{\pi}{2}-1) B(\alpha,(k,l,c',d'))$ to the event $(k,l,0,0)$ in the full model $\mathcal{M}$. They result from the submodels  $\mathcal{M}_{(k,l,c',d')}$, each drawn with the probability \eqref{probLarsson}, giving constant probabilities $\frac{1}{4}- \frac{1}{2\pi}$ for the events from $\mathcal O$.
The sum of all such contributions cannot be greater than the quantum probability for the event, $p(k,l,0,0)$, but can be lower since the difference can be compensated by the trivial models. This gives the following consistency conditions
\begin{eqnarray}
\label{condition}
\Delta_{(k,l,0,0)}=
p(k,l,0,0)\nonumber \\
-\left(\frac{\pi}{2}-1\right)\sum\limits_{c'> d'} B(\alpha,(k,l,c',d'))\geq 0,
\end{eqnarray}
which must hold for any $k\neq l$.
Obviously, due to the symmetrization an analogous condition can be written for events of $(0,0,r,s)$ type.

In the Appendix C we show that the condition in Eq. (\ref{condition}) is satisfied for  
{\em any} $(k,l)$ and $(r,s)$, whenever $\alpha^2<0.87$. The threshold value is given by the case $|k-l|=1$, as the larger this difference, the higher the $\alpha$ for which the condition holds. 

The model can be completed using a family of trivial submodels $\mathcal{M}_{\mathbf{n}}$ for events $\mathbf{n}\in\mathcal{N}$. They predict fixed outcomes for Alice and Bob, $P^A_{\mathbf{n}}(k,l)=P^B_{\mathbf{n}}(r,s)=1$, which lead to $P^{AB}_\mathbf{n}((k,l,r,s))=1$. 
Obviously, for events $\mathbf{n}\in\mathcal{N}\setminus{\mathcal{O}}$, we choose each corresponding trivial model $\mathcal{M_{\mathbf{n}}}$ with probability $p(\mathbf{n})$. Finally, for events $\mathbf{n}\in\mathcal{O}$ we might need to
compensate the potential difference $\Delta_{(k,l,0,0)} > 0$ between the quantum predictions for the $\mathcal{O}$-events and the predictions specified by the Larsson-like models. To do that, we use an additional  trivial submodel for event ${(k,l,0,0)}$, which appears in the full model  with probability $P(\mathcal{M}_{(k,l,0,0)})=\Delta_{(k,l,0,0)}$.
The case of $\Delta_{(0,0,r,s)}>0$ is treated the same way. 
 
One can easily build a better version of the model which would hold for slightly higher values of $\alpha$. However, we were not able to find a model which has an unconstrained validity, and one can 
conjecture that the Larsson-like approach cannot lead to such.  Still, our model fully covers the range of $\alpha$ for which TWC
predicted a violation of local realism. Thus, this claim is fully revoked.

{\it Closing remarks.} The above result closes the ambiguity of the relation of TWC correlations with violations of local realism.  Similar form of correlations for coincident counts stemmed from several other experimental schemes proposed in the early times of photonic entanglement-interferometry. Their functional sine-like  dependence $v\sin{(\theta_1 \pm \theta_2)}$ was considered to violate a Bell inequality when the
visibility $v$, was higher than $\frac{1}{\sqrt{2}}$. Such dependence appears, for instance, in the standard two qubit Bell experiments, and $v>\frac{1}{\sqrt{2}}$ indeed blocks any possibility of having a local realistic model in the idealized scenario (perfect efficiency, etc.). The most important
claims of this character were put in \cite{Mandel88}, for a different situation in \cite{Franson89}, and in a still different experimental context in the TWC paper. Experiments of \cite{Mandel88} were shown to be, in their idealized versions, proper Bell experiments \cite{PSHZ97}. However Franson interferometry, even in the ideal form, was shown to have a local realistic model reproducing the correlations for all values of the local phases \cite{Franson99}. 
Modifications of the Franson scheme were shown to be a necessity. Here we showed that the idealized TWC correlations have also a simple model in the region in which they were supposed to violate local realism. One has to modify the experiment in order to see a genuine violation of local realism, e.g., like in the proposal of \cite{Banaszek99} or \cite{Hardy94}. However this is not a minor modification. In a forthcoming paper we shall analyse 
modifications which have some of the traits of the one of \cite{Hardy94}, but still involve just initial photon in mode $s$, Fig.(\ref{mainSetup}), like it is in the case of \cite{TWC91}. They cannot have a local realistic model. 

We stress that the results so far presented have a twofold value. We answer the \emph{vexata quaestio} about the `nonlocality of a single photon' as presented in \cite{TWC91}, but we also put a warning on its possible exploitation for future quantum technologies application. 
In fact, the latter is a rising field with a flourishing literature and our findings should serve as a \emph{caveat lector} for any attempt to use this scheme in any protocol requiring as the main resource the violation of local realism.

\section*{Acknowledgements}
We acknowledge support by the Foundation for Polish Science (IRAP project, ICTQT, contract no. 2018/MAB/5, co-financed by EU within Smart Growth Operational Programme). MK acknowledges support by the Foundation for Polish Science through the START scholarship. 
AM acknowledges support by National Research Center through the grant MINIATURA  DEC-2020/04/X/ST2/01794.

\bibliography{SinglePhoton}

\begin{thebibliography}{16}%
\makeatletter
\providecommand \@ifxundefined [1]{%
 \@ifx{#1\undefined}
}%
\providecommand \@ifnum [1]{%
 \ifnum #1\expandafter \@firstoftwo
 \else \expandafter \@secondoftwo
 \fi
}%
\providecommand \@ifx [1]{%
 \ifx #1\expandafter \@firstoftwo
 \else \expandafter \@secondoftwo
 \fi
}%
\providecommand \natexlab [1]{#1}%
\providecommand \enquote  [1]{``#1''}%
\providecommand \bibnamefont  [1]{#1}%
\providecommand \bibfnamefont [1]{#1}%
\providecommand \citenamefont [1]{#1}%
\providecommand \href@noop [0]{\@secondoftwo}%
\providecommand \href [0]{\begingroup \@sanitize@url \@href}%
\providecommand \@href[1]{\@@startlink{#1}\@@href}%
\providecommand \@@href[1]{\endgroup#1\@@endlink}%
\providecommand \@sanitize@url [0]{\catcode `\\12\catcode `\$12\catcode
  `\&12\catcode `\#12\catcode `\^12\catcode `\_12\catcode `\%12\relax}%
\providecommand \@@startlink[1]{}%
\providecommand \@@endlink[0]{}%
\providecommand \url  [0]{\begingroup\@sanitize@url \@url }%
\providecommand \@url [1]{\endgroup\@href {#1}{\urlprefix }}%
\providecommand \urlprefix  [0]{URL }%
\providecommand \Eprint [0]{\href }%
\providecommand \doibase [0]{https://doi.org/}%
\providecommand \selectlanguage [0]{\@gobble}%
\providecommand \bibinfo  [0]{\@secondoftwo}%
\providecommand \bibfield  [0]{\@secondoftwo}%
\providecommand \translation [1]{[#1]}%
\providecommand \BibitemOpen [0]{}%
\providecommand \bibitemStop [0]{}%
\providecommand \bibitemNoStop [0]{.\EOS\space}%
\providecommand \EOS [0]{\spacefactor3000\relax}%
\providecommand \BibitemShut  [1]{\csname bibitem#1\endcsname}%
\let\auto@bib@innerbib\@empty
\bibitem [{\citenamefont {Tan}\ \emph {et~al.}(1991)\citenamefont {Tan},
  \citenamefont {Walls},\ and\ \citenamefont {Collett}}]{TWC91}%
  \BibitemOpen
  \bibfield  {author} {\bibinfo {author} {\bibfnamefont {S.~M.}\ \bibnamefont
  {Tan}}, \bibinfo {author} {\bibfnamefont {D.~F.}\ \bibnamefont {Walls}},\
  and\ \bibinfo {author} {\bibfnamefont {M.~J.}\ \bibnamefont {Collett}},\
  }\bibfield  {title} {\bibinfo {title} {Nonlocality of a single photon},\
  }\href {https://doi.org/10.1103/PhysRevLett.66.252} {\bibfield  {journal}
  {\bibinfo  {journal} {Phys. Rev. Lett.}\ }\textbf {\bibinfo {volume} {66}},\
  \bibinfo {pages} {252} (\bibinfo {year} {1991})}\BibitemShut {NoStop}%
\bibitem [{\citenamefont {Reid}\ and\ \citenamefont {Walls}(1986)}]{Reid86}%
  \BibitemOpen
  \bibfield  {author} {\bibinfo {author} {\bibfnamefont {M.~D.}\ \bibnamefont
  {Reid}}\ and\ \bibinfo {author} {\bibfnamefont {D.~F.}\ \bibnamefont
  {Walls}},\ }\bibfield  {title} {\bibinfo {title} {Violations of classical
  inequalities in quantum optics},\ }\href
  {https://doi.org/10.1103/PhysRevA.34.1260} {\bibfield  {journal} {\bibinfo
  {journal} {Phys. Rev. A}\ }\textbf {\bibinfo {volume} {34}},\ \bibinfo
  {pages} {1260} (\bibinfo {year} {1986})}\BibitemShut {NoStop}%
\bibitem [{\citenamefont {\ifmmode~\dot{Z}\else \.{Z}\fi{}ukowski}\ \emph
  {et~al.}(2016)\citenamefont {\ifmmode~\dot{Z}\else \.{Z}\fi{}ukowski},
  \citenamefont {Wie\ifmmode~\acute{s}\else \'{s}\fi{}niak},\ and\
  \citenamefont {Laskowski}}]{Zukowski16}%
  \BibitemOpen
  \bibfield  {author} {\bibinfo {author} {\bibfnamefont {M.}~\bibnamefont
  {\ifmmode~\dot{Z}\else \.{Z}\fi{}ukowski}}, \bibinfo {author} {\bibfnamefont
  {M.}~\bibnamefont {Wie\ifmmode~\acute{s}\else \'{s}\fi{}niak}},\ and\
  \bibinfo {author} {\bibfnamefont {W.}~\bibnamefont {Laskowski}},\ }\bibfield
  {title} {\bibinfo {title} {Bell inequalities for quantum optical fields},\
  }\href {https://doi.org/10.1103/PhysRevA.94.020102} {\bibfield  {journal}
  {\bibinfo  {journal} {Phys. Rev. A}\ }\textbf {\bibinfo {volume} {94}},\
  \bibinfo {pages} {020102} (\bibinfo {year} {2016})}\BibitemShut {NoStop}%
\bibitem [{\citenamefont {Santos}(1992)}]{Santos92}%
  \BibitemOpen
  \bibfield  {author} {\bibinfo {author} {\bibfnamefont {E.}~\bibnamefont
  {Santos}},\ }\bibfield  {title} {\bibinfo {title} {Comment on ``nonlocality
  of a single photon''},\ }\href {https://doi.org/10.1103/PhysRevLett.68.894}
  {\bibfield  {journal} {\bibinfo  {journal} {Phys. Rev. Lett.}\ }\textbf
  {\bibinfo {volume} {68}},\ \bibinfo {pages} {894} (\bibinfo {year}
  {1992})}\BibitemShut {NoStop}%
\bibitem [{\citenamefont {Greenberger}\ \emph {et~al.}(1995)\citenamefont
  {Greenberger}, \citenamefont {Horne},\ and\ \citenamefont
  {Zeilinger}}]{Greenberger95}%
  \BibitemOpen
  \bibfield  {author} {\bibinfo {author} {\bibfnamefont {D.~M.}\ \bibnamefont
  {Greenberger}}, \bibinfo {author} {\bibfnamefont {M.~A.}\ \bibnamefont
  {Horne}},\ and\ \bibinfo {author} {\bibfnamefont {A.}~\bibnamefont
  {Zeilinger}},\ }\bibfield  {title} {\bibinfo {title} {Nonlocality of a single
  photon?},\ }\href {https://doi.org/10.1103/PhysRevLett.75.2064} {\bibfield
  {journal} {\bibinfo  {journal} {Phys. Rev. Lett.}\ }\textbf {\bibinfo
  {volume} {75}},\ \bibinfo {pages} {2064} (\bibinfo {year}
  {1995})}\BibitemShut {NoStop}%
\bibitem [{\citenamefont {Peres}(1995)}]{Peres95}%
  \BibitemOpen
  \bibfield  {author} {\bibinfo {author} {\bibfnamefont {A.}~\bibnamefont
  {Peres}},\ }\bibfield  {title} {\bibinfo {title} {Nonlocal effects in {Fock}
  space},\ }\href {https://doi.org/10.1103/PhysRevLett.74.4571} {\bibfield
  {journal} {\bibinfo  {journal} {Phys. Rev. Lett.}\ }\textbf {\bibinfo
  {volume} {74}},\ \bibinfo {pages} {4571} (\bibinfo {year}
  {1995})}\BibitemShut {NoStop}%
\bibitem [{\citenamefont {Hardy}(1994)}]{Hardy94}%
  \BibitemOpen
  \bibfield  {author} {\bibinfo {author} {\bibfnamefont {L.}~\bibnamefont
  {Hardy}},\ }\bibfield  {title} {\bibinfo {title} {Nonlocality of a single
  photon revisited},\ }\href {https://doi.org/10.1103/PhysRevLett.73.2279}
  {\bibfield  {journal} {\bibinfo  {journal} {Phys. Rev. Lett.}\ }\textbf
  {\bibinfo {volume} {73}},\ \bibinfo {pages} {2279} (\bibinfo {year}
  {1994})}\BibitemShut {NoStop}%
\bibitem [{\citenamefont {Banaszek}\ and\ \citenamefont
  {W\'odkiewicz}(1999)}]{Banaszek99}%
  \BibitemOpen
  \bibfield  {author} {\bibinfo {author} {\bibfnamefont {K.}~\bibnamefont
  {Banaszek}}\ and\ \bibinfo {author} {\bibfnamefont {K.}~\bibnamefont
  {W\'odkiewicz}},\ }\bibfield  {title} {\bibinfo {title} {Testing quantum
  nonlocality in phase space},\ }\href
  {https://doi.org/10.1103/PhysRevLett.82.2009} {\bibfield  {journal} {\bibinfo
   {journal} {Phys. Rev. Lett.}\ }\textbf {\bibinfo {volume} {82}},\ \bibinfo
  {pages} {2009} (\bibinfo {year} {1999})}\BibitemShut {NoStop}%
\bibitem [{\citenamefont {van Enk}(2005)}]{Enk05}%
  \BibitemOpen
  \bibfield  {author} {\bibinfo {author} {\bibfnamefont {S.~J.}\ \bibnamefont
  {van Enk}},\ }\bibfield  {title} {\bibinfo {title} {Single-particle
  entanglement},\ }\href {https://doi.org/10.1103/PhysRevA.72.064306}
  {\bibfield  {journal} {\bibinfo  {journal} {Phys. Rev. A}\ }\textbf {\bibinfo
  {volume} {72}},\ \bibinfo {pages} {064306} (\bibinfo {year}
  {2005})}\BibitemShut {NoStop}%
\bibitem [{\citenamefont {Hessmo}\ \emph {et~al.}(2004)\citenamefont {Hessmo},
  \citenamefont {Usachev}, \citenamefont {Heydari},\ and\ \citenamefont
  {Bj\"ork}}]{Hessmo04}%
  \BibitemOpen
  \bibfield  {author} {\bibinfo {author} {\bibfnamefont {B.}~\bibnamefont
  {Hessmo}}, \bibinfo {author} {\bibfnamefont {P.}~\bibnamefont {Usachev}},
  \bibinfo {author} {\bibfnamefont {H.}~\bibnamefont {Heydari}},\ and\ \bibinfo
  {author} {\bibfnamefont {G.}~\bibnamefont {Bj\"ork}},\ }\bibfield  {title}
  {\bibinfo {title} {Experimental demonstration of single photon nonlocality},\
  }\href {https://doi.org/10.1103/PhysRevLett.92.180401} {\bibfield  {journal}
  {\bibinfo  {journal} {Phys. Rev. Lett.}\ }\textbf {\bibinfo {volume} {92}},\
  \bibinfo {pages} {180401} (\bibinfo {year} {2004})}\BibitemShut {NoStop}%
\bibitem [{\citenamefont {Babichev}\ \emph {et~al.}(2004)\citenamefont
  {Babichev}, \citenamefont {Appel},\ and\ \citenamefont
  {Lvovsky}}]{Babichev04}%
  \BibitemOpen
  \bibfield  {author} {\bibinfo {author} {\bibfnamefont {S.~A.}\ \bibnamefont
  {Babichev}}, \bibinfo {author} {\bibfnamefont {J.}~\bibnamefont {Appel}},\
  and\ \bibinfo {author} {\bibfnamefont {A.~I.}\ \bibnamefont {Lvovsky}},\
  }\bibfield  {title} {\bibinfo {title} {Homodyne tomography characterization
  and nonlocality of a dual-mode optical qubit},\ }\href
  {https://doi.org/10.1103/PhysRevLett.92.193601} {\bibfield  {journal}
  {\bibinfo  {journal} {Phys. Rev. Lett.}\ }\textbf {\bibinfo {volume} {92}},\
  \bibinfo {pages} {193601} (\bibinfo {year} {2004})}\BibitemShut {NoStop}%
\bibitem [{\citenamefont {Åke Larsson}(1999)}]{LARSSON99}%
  \BibitemOpen
  \bibfield  {author} {\bibinfo {author} {\bibfnamefont {J.}~\bibnamefont {Åke
  Larsson}},\ }\bibfield  {title} {\bibinfo {title} {Modeling the singlet state
  with local variables},\ }\href
  {https://doi.org/https://doi.org/10.1016/S0375-9601(99)00236-4} {\bibfield
  {journal} {\bibinfo  {journal} {Physics Letters A}\ }\textbf {\bibinfo
  {volume} {256}},\ \bibinfo {pages} {245 } (\bibinfo {year}
  {1999})}\BibitemShut {NoStop}%
\bibitem [{\citenamefont {Aerts}\ \emph {et~al.}(1999)\citenamefont {Aerts},
  \citenamefont {Kwiat}, \citenamefont {Larsson},\ and\ \citenamefont
  {\.Zukowski}}]{Franson99}%
  \BibitemOpen
  \bibfield  {author} {\bibinfo {author} {\bibfnamefont {S.}~\bibnamefont
  {Aerts}}, \bibinfo {author} {\bibfnamefont {P.}~\bibnamefont {Kwiat}},
  \bibinfo {author} {\bibfnamefont {J.-A.}\ \bibnamefont {Larsson}},\ and\
  \bibinfo {author} {\bibfnamefont {M.}~\bibnamefont {\.Zukowski}},\ }\bibfield
   {title} {\bibinfo {title} {Two-photon {Franson}-type experiments and local
  realism},\ }\href {https://doi.org/10.1103/PhysRevLett.83.2872} {\bibfield
  {journal} {\bibinfo  {journal} {Phys. Rev. Lett.}\ }\textbf {\bibinfo
  {volume} {83}},\ \bibinfo {pages} {2872} (\bibinfo {year}
  {1999})}\BibitemShut {NoStop}%
\bibitem [{\citenamefont {Ou}\ and\ \citenamefont {Mandel}(1988)}]{Mandel88}%
  \BibitemOpen
  \bibfield  {author} {\bibinfo {author} {\bibfnamefont {Z.~Y.}\ \bibnamefont
  {Ou}}\ and\ \bibinfo {author} {\bibfnamefont {L.}~\bibnamefont {Mandel}},\
  }\bibfield  {title} {\bibinfo {title} {Violation of {Bell's} inequality and
  classical probability in a two-photon correlation experiment},\ }\href
  {https://doi.org/10.1103/PhysRevLett.61.50} {\bibfield  {journal} {\bibinfo
  {journal} {Phys. Rev. Lett.}\ }\textbf {\bibinfo {volume} {61}},\ \bibinfo
  {pages} {50} (\bibinfo {year} {1988})}\BibitemShut {NoStop}%
\bibitem [{\citenamefont {Franson}(1989)}]{Franson89}%
  \BibitemOpen
  \bibfield  {author} {\bibinfo {author} {\bibfnamefont {J.~D.}\ \bibnamefont
  {Franson}},\ }\bibfield  {title} {\bibinfo {title} {{Bell} inequality for
  position and time},\ }\href {https://doi.org/10.1103/PhysRevLett.62.2205}
  {\bibfield  {journal} {\bibinfo  {journal} {Phys. Rev. Lett.}\ }\textbf
  {\bibinfo {volume} {62}},\ \bibinfo {pages} {2205} (\bibinfo {year}
  {1989})}\BibitemShut {NoStop}%
\bibitem [{\citenamefont {Popescu}\ \emph {et~al.}(1997)\citenamefont
  {Popescu}, \citenamefont {Hardy},\ and\ \citenamefont {\ifmmode~\dot{Z}\else
  \.{Z}\fi{}ukowski}}]{PSHZ97}%
  \BibitemOpen
  \bibfield  {author} {\bibinfo {author} {\bibfnamefont {S.}~\bibnamefont
  {Popescu}}, \bibinfo {author} {\bibfnamefont {L.}~\bibnamefont {Hardy}},\
  and\ \bibinfo {author} {\bibfnamefont {M.}~\bibnamefont
  {\ifmmode~\dot{Z}\else \.{Z}\fi{}ukowski}},\ }\bibfield  {title} {\bibinfo
  {title} {Revisiting {Bell's} theorem for a class of down-conversion
  experiments},\ }\href {https://doi.org/10.1103/PhysRevA.56.R4353} {\bibfield
  {journal} {\bibinfo  {journal} {Phys. Rev. A}\ }\textbf {\bibinfo {volume}
  {56}},\ \bibinfo {pages} {R4353} (\bibinfo {year} {1997})}\BibitemShut
  {NoStop}%
\end{thebibliography}%
 
\onecolumngrid

\appendix 

\section*{Appendix A -- Quantum photodetection probabilities}
In this section, we are going to calculate the probability of detecting the event $n=(k_{c_1},l_{d_1},r_{c_2},s_{d_2})$, consisting in registering  specific numbers of photons in the output modes of the Tan-Walls-Collett setup.

The initial state, obtained by transforming a single photon with a balanced beamsplitter and adding two coherent states as ancillas reads
\begin{equation}
\label{a1}
|\Psi\rangle=|\alpha e^{i \theta_1}\rangle_{a_1}  \frac{1}{\sqrt{2}} (|01\rangle+i\,|10\rangle)_{b_1b_2}|\alpha e^{i \theta_2}\rangle_{a_2}.
\end{equation}

We will show how the state \eqref{a1} transforms on balanced beamsplitters $U_{BSj}, ~j = 1,2$ which link the output and input modes via
\begin{equation}
\label{a2}
   \hat c_j=\frac{1}{\sqrt2}(\hat a_j+i \, \hat b_j) \,\, \text{and} \, \,  \hat d_j=\frac{1}{\sqrt2}(i\,\hat a_j+ \hat b_j).
\end{equation}
Applying \eqref{a2} to the state \eqref{a1} we get
\begin{eqnarray}
\ket{\Psi} &=& e^{-\alpha^2 } \sum_{j=0}^{\infty}  \frac{(\alpha e^{i \theta_1})^j}{j!} (\hat a_1^\dagger)^j \frac{1}{\sqrt{2}} ( i\hat b_1^\dagger  + \hat b_2^{\dagger})   \sum_{k=0}^{\infty}  \frac{(\alpha e^{i \theta_2})^k}{k!} (\hat a_2^\dagger)^k \nonumber\\
&=& e^{-\alpha^2 }\sum_{j,k=0}^{\infty} {2}^{-\frac{j+k}{2}}\frac{(\alpha e^{i \theta_1})^j}{{j!}}\frac{(\alpha e^{i \theta_2})^k}{{k!}}\Big( \hat c_1^{\dagger} +i\hat d^{\dagger}_1 \Big)^j\frac 12 \Big(-\hat c_1^{\dagger} + i\hat d_1^\dagger  +i \hat c_2^\dagger + \hat d_2^{\dagger}\Big)  \Big( \hat c_2^{\dagger} + i\hat d^{\dagger}_2 \Big)^k\ket{\Omega} \nonumber \\
&=&  e^{-\alpha^2 }\sum_{j,k=0}^{\infty} {2}^{-\frac{j+k}{2}}\frac{(\alpha e^{i \theta_1})^j}{{j!}}\frac{(\alpha e^{i \theta_2})^k}{{k!}}\sum_{p=0}^j\binom{j}{p}(\hat c^{\dagger}_1)^{j-p}(i\hat d_1^{\dagger})^{p}\frac 12 \Big(-\hat c_1^{\dagger} + i\hat d_1^\dagger  +i \hat c_2^\dagger + \hat d_2^{\dagger}\Big)  \sum_{q=0}^k\binom{k}{q}(\hat c^{\dagger}_2)^{k-q}(i\hat d_2^{\dagger})^{q} \ket{\Omega},\nonumber \\
&=& \sum_{j,k=0}^{\infty}  \sum_{p=0}^j \sum_{q=0}^k f(j,p,k,q) (\hat c^{\dagger}_1)^{j-p}(\hat d_1^{\dagger})^{p} \Big(-\hat c_1^{\dagger} + i\hat d_1^\dagger  +i \hat c_2^\dagger + \hat d_2^{\dagger}\Big)(\hat c^{\dagger}_2)^{k-q}(\hat d_2^{\dagger})^{q} \ket{\Omega}, \\
&=&  \sum_{j,k=0}^{\infty}\sum_{p}^{j}\sum_{q=0}^kf(j,p,k,q)\bigg[ - \sqrt{(j-p+1)!p!(k-q)!q!}\ket{j-p+1}_{c_1}\ket{p}_{d_1}\ket{k-q}_{c_2}\ket{q}_{d_2}  \nonumber \\
&& + i\sqrt{(j-p)!(p+1)!(k-q)!q!}\ket{j-p}_{c_1}\ket{p+1}_{d_1}\ket{k-q}_{c_2}\ket{q}_{d_2}\nonumber \\
&& + i\sqrt{(j-p)!p!(k-q+1)!q!}\ket{j-p}_{c_1}\ket{p}_{d_1}\ket{k-q+1}_{c_2}\ket{q}_{d_2} \nonumber \\
&& +  \sqrt{(j-p)!p!(k-q)!(q+1)!}\ket{j-p}_{c_1}\ket{p}_{d_1}\ket{k-q}_{c_2}\ket{q+1}_{d_2}  
\bigg] 
\end{eqnarray}
where 
\begin{eqnarray}
f(j,p,k,q) &=& e^{-\alpha^2 } {2}^{-\frac{j+k}{2}-1} \frac{(\alpha e^{i \theta_1})^j}{{j!}}\frac{(\alpha e^{i \theta_2})^k}{{k!}} \binom{j}{p} \binom{k}{q}  (i)^{p+q}, ~~~~ \forall p \leq j, q \leq k.
\label{gFUNCTION}
\end{eqnarray} 
Now, 
\begin{eqnarray} 
&& \text{Pr}(k, l; r, s) = |\bra{k,l,r,s}\ket{\Psi}|^2 \nonumber \\
&=& \bigg|
- f(k+l-1,l,r+s,s) + i f(k+l-1,l - 1,r+s,s)  \nonumber \\
&& \hspace{4cm} +i f(k+l,l,r+s-1,s)
+f(k+l,l,r+s-1,s-1)\bigg|^2  k!~ l!~ r! ~ s! \nonumber \\
&=& \frac{e^{-2\alpha^2}}{ k!~ l!~ r! ~ s!} \Big(\frac{\alpha^2}{2}\Big)^{k+l+r+s} \frac{1}{2\alpha^2}  \Big[ (k-l)^2 + (r-s)^2 + 2(k-l)(r-s) \sin(\theta_1 - \theta_2) \Big], ~~
\label{eqapp:probability}
\end{eqnarray} 

\section*{Appendix B -- On the sum of probabilities of all submodels $\mathcal{M_\mathbf{n}}$}

In this section we prove that the probabilities $P(\mathcal{M_\mathbf{n}})$ of choosing specific submodels are properly normalized. We have
\begin{eqnarray}
\label{b1}
    \sum\limits_{\mathbf{n}\in\mathcal{N}\cap\mathcal{\tilde{N}}}P(\mathcal{M_\mathbf{n}})=\sum\limits_{\mathbf{n}\in\mathcal{N}\setminus\mathcal{O}}B(\alpha,\mathcal{M_\mathbf{n}})+\sum\limits_{\mathbf{n}\in\mathcal{\tilde{N}}}2\pi B(\alpha,\mathcal{M_\mathbf{n}})+\sum\limits_{\mathbf{n}\in\mathcal{O}}\Delta_\mathbf{n},
\end{eqnarray}
where

\begin{eqnarray}
\label{b2}
  \sum\limits_{\mathbf{n}\in\mathcal{O}}\Delta_\mathbf{n}=
\sum\limits_{k\neq l}  \left(p(\mathbf{(k,l,0,0)})-\left(\frac{\pi}{2}-1\right)\sum\limits_{c'> d'} B(\alpha,(k,l,c',d'))\right)+
\sum\limits_{r\neq s}  \left(p(\mathbf{(0,0,r,s)})-\left(\frac{\pi}{2}-1\right)\sum\limits_{c> d} B(\alpha,(c,d,r,s))\right)\nonumber\\
=\sum\limits_{\mathbf{n}\in\mathcal{O}}B(\alpha,\mathcal{M_\mathbf{n}})-4\left(\frac{\pi}{2}-1\right)\sum\limits_{\mathbf{n}\in\mathcal{\tilde{N}}} B(\alpha,\mathcal{M_\mathbf{n}})=\sum\limits_{\mathbf{n}\in\mathcal{O}}B(\alpha,\mathcal{M_\mathbf{n}})-\left(2\pi-4\right)\sum\limits_{\mathbf{n}\in\mathcal{\tilde{N}}} B(\alpha,\mathcal{M_\mathbf{n}}).
\end{eqnarray}
Moreover, notice that
\begin{equation}
\label{b3}
    \sum\limits_{\mathbf{n}\in\mathcal{\tilde{N}}} B(\alpha,\mathcal{M_\mathbf{n}})=\frac{1}{4} \sum\limits_{\substack{\mathbf{n}\in\mathbb{N^4}\\c\neq d,\, c'\neq d'}} B(\alpha,\mathcal{M_\mathbf{n}}).
\end{equation}
Plugging Eqs.(\ref{b2}) and (\ref{b3} )into Eq. (\ref{b1}) we get
\begin{eqnarray}
    &\sum\limits_{\mathbf{n}\in\mathcal{N}\cap\mathcal{\tilde{N}}}P(\mathcal{M_\mathbf{n}})=\sum\limits_{\mathbf{n}\in\mathcal{N}\setminus\mathcal{O}}B(\alpha,\mathcal{M_\mathbf{n}})+2\pi\sum\limits_{\mathbf{n}\in\mathcal{\tilde{N}}} B(\alpha,\mathcal{M_\mathbf{n}}) \nonumber \\
    &+\sum\limits_{\mathbf{n}\in\mathcal{O}}B(\alpha,\mathcal{M_\mathbf{n}})-\left(2\pi-4\right)\sum\limits_{\mathbf{n}\in\mathcal{\tilde{N}}} B(\alpha,\mathbf{n})=\sum_{\mathbf{n}\in\mathbb{N^4}}B(\alpha,\mathbf{n})=1.
\end{eqnarray}

\section*{Appendix C -- Threshold intensity for the validity of the LHV model}

In this section we prove that if $\alpha^2<0.87$, the probabilities of choosing a specific submodel $ P(\mathcal{M_\mathbf{n}})$ are non-negative. To do that, we only need to consider $\mathbf{n_0}\in\mathcal{O}$, for which $P(\mathcal{M}_\mathbf{n_0})=\Delta_\mathbf{n_0}$. Let us fix $\mathbf{n_0}=(k,l,0,0), \,k\neq l$, as the reasoning for $\mathbf{n_0}=(0,0,r,s)$ is fully analogous. We need to check the conditions in which 
\begin{eqnarray}
\label{c1}
\Delta_\mathbf{n_0}=B(\alpha,\mathbf{(k,l,0,0)})-\left(\frac{\pi}{2}-1\right)\sum\limits_{c'> d'} B(\alpha,(k,l,c',d')\geq 0.
\end{eqnarray}
We plug the definition of the function $B(\alpha,\mathbf{n})$ from the main text into \eqref{c1} and obtain, after some transformations,
\begin{eqnarray}
\Delta_\mathbf{n_0}=\frac{e^{-2 \alpha^2} 2^{-k-l-3} \left(\alpha^2\right)^{k+l-1} \left(-(\pi -2) e^{\alpha^2} \left(\alpha^2+(k-l)^2\right)+(\pi -2) I_0\left(\alpha^2\right) (k-l)^2+4 (k-l)^2\right)}{k!l!}.
\end{eqnarray}
It is easy to see that the condition $\Delta_\mathbf{n_0}\geq0$ is equivalent to
\begin{equation}
\label{c2}
    -(\pi -2) e^{\alpha^2} \left(\alpha^2+(k-l)^2\right)+(\pi -2) I_0\left(\alpha^2\right) (k-l)^2+4 (k-l)^2\geq0.
\end{equation}
As the Bessel function $I_0$ satisfies  $I_0\left(\alpha^2\right)\geq1$, the inequality \eqref{c2} can be approximated by a slightly stricter
\begin{equation}
    -(\pi -2) e^{\alpha^2} \left(\alpha^2+(k-l)^2\right)+(\pi -2)(k-l)^2+4 (k-l)^2=\left((\pi -2) \left(-e^{\alpha^2}\right)+\pi +2\right) (k-l)^2-(\pi -2) \alpha^2 e^{\alpha^2}\geq0.
\end{equation}
For $\alpha<1$, the coefficient $\left((\pi -2) \left(-e^{\alpha^2}\right)+\pi +2\right)$ standing in front of $(k-l)^2$ is positive. This means that the critical case we need to consider is $(k-l)^2=1$. Thus, we arrive at
\begin{equation}
\label{c3}
   \left((\pi -2) \left(-e^{\alpha^2}\right)+\pi +2\right) -(\pi -2) \alpha^2 e^{\alpha^2}\geq0.
\end{equation}
It can be shown that the inequality \ref{c3} is satisfied for \begin{equation}
    \alpha^2\leq W\left(\frac{2 e+e \pi }{\pi -2}\right)-1\approx 0.87,
\end{equation}
where $W$ denotes the Lambert $W$ function ($W(z)$ returns the principal solution for $w$ in $z=w e^w$ ).


\end{document}